\documentclass[12pt]{article}
\usepackage{amsmath,amssymb}
\usepackage{graphicx}
\usepackage[pdftex]{hyperref}
\DeclareGraphicsExtensions{.eps,.bmp,.wmf,.jpeg,.pdf}
\numberwithin{equation}{section}
\def\be{\begin{equation}}
\def\ee{\end{equation}}

\def\bea{\begin{eqnarray}}
\def\eea{\end{eqnarray}}

\title{Inflation driven by scalar field with non-minimal kinetic coupling with Higgs and quadratic potentials}
\author{L.N. Granda\thanks{ngranda@univalle.edu.co} \\{\small\it Departamento de Fisica, Universidad del Valle}\\{\small\it A.A. 25360, Cali, Colombia}\\{\small\it and}\\
{\small\it Departamento de Fisica, Universidad de Murcia}\\{\small\it 30100 Murcia, Spain}}
\date{}
\begin{document}
\maketitle

\begin{abstract}
\noindent We study a scalar field with non-minimal kinetic coupling to itself and to the curvature. The slow rolling conditions allowing an inflationary background have been found. The quadratic and Higgs type potentials have been considered, and the corresponding values for the scalar fields at the end of inflation allows to recover the connection with particle physics. \\

%\noindent \textbf{Keywords:}  Non-minimal kinetic, inflation\\
%\noindent PACS 98.80.-k, 98.80.Cq, 04.50.kd
\end{abstract}

\section{Introduction}
\noindent 
The latest astrophysical observations \cite{komatsu} account for the fact that the universe is homogeneous, isotropic and spatially flat, at large scales, i.e. at cosmological scales, the observable
universe looks surprisingly smooth, except for a tiny fluctuations. According to this, the universe is well described by a Friedmann-Robertson-Walker (FRW)
spatially flat geometry, which depends on a single scale factor. The the homogeneity and isotropy imply that all points swept by the CMB observations should have been
in causal contact, with a high degree of accuracy, at very early times. Besides this, there is the problem of the current spatial flatness, which requires an
extremely fine tuning of the spatial curvature at very early times. These problems can not be solved in the frame of the Einstein gravity with the standard model particle physics as the only source.
The most studied solution to the above problems lies in the phenomenon known as inflation \cite {guth}, \cite{linde}, \cite{steinhardt}, \cite{linde1}, \cite{lyth}. The inflation is seen as a rapid expansion of the early time universe, which solves the homogeneity, isotropy and flatness problems. During this inflation era, the universe was filled with essentially a homogeneous scalar field, called
the inflaton, with a potential dominating the energy density of the universe, and decreasing slowly with time as the scalar field rolled slowly down the slope of the potential. Among others, one of the virtues of inflation is that it can set the initial conditions for the subsequent hot big bang, which otherwise have to be imposed. By other hand, fundamental scalar fields are not yet discovered in nature, but they appear in different
extensions of the standard model, involving grand unified theories, supersymmetry, string theory, higher dimensional gravity theories, etc. Inflaton properties are constrained by the
observations of fluctuations of the Cosmic Microwave Background (CMB) and the matter distribution in the universe. So in principle the hypothesis of inflation represents a theoretical problem that still to be solved. Initially, the inflation has been realized by a minimally coupled scalar field \cite{linde2}, but the hypothesis of inflation might well be realized by many other models like kinetic inflation \cite{picon}; vector inflation \cite{ford}, \cite{mota}, \cite{golovnev}; inflaton potential in supergravity \cite{kawasaki}, \cite{davis}, \cite{linde3}; string theory inspired inflation \cite{kallosh}, to mention some of them. \\ 
In the present work, we concentrate in a class of models with non-minimal coupling to gravity, specifically in a scalar field model whose kinetic term is non-minimally coupled to gravity and to the scalar field \cite{granda}. This kind of couplings appear as low energy limit of several higher dimensional theories (see section $9.5$ in \cite{linde1}), and provide a possible approach to quantum gravity from a perturbative point of view \cite{donoghue}. A coupling between curvature and kinetic terms also appears as part of the Weyl anomaly in $N=4$ conformal supergravity (i.e. as a part of the action of the $N=4$ conformal supergravity) \cite{tseytlin, odintsov2}, as quantum corrections to Brans-Dicke theory \cite{elizalde} and in different frames in modified gravity \cite{nojiri}. A model with non-minimal derivative couplings was proposed in \cite{amendola2}, \cite{capozziello1}, \cite{capozziello2} in the context of inflationary cosmology, and recently, non-minimal derivative coupling of the Higgs field was considered in \cite{germani}, also as inflationary model. In \cite{caldwell1} a derivative coupling to Ricci tensor has been considered to study cosmological restrictions on the coupling parameter, and the role of this coupling during inflation. Some asymptotic solutions for a non-minimal kinetic coupling to scalar and Ricci curvatures were found in \cite{sushkov}, and quintessence and phantom cosmological scenarios with non-minimal derivative coupling have been studied in \cite{saridakis}. The cosmological dynamics of scalar field with kinetic term coupled to a product of Einstein tensors, has been analyzed in \cite{gao}. An inflationary model with non-minimally coupled inflaton and Dirac-Born-Infeld (DBI) kinetic term, has been proposed in \cite{easson3}. In the context of modified $f(R)$ gravity theories, a class of $f(R)$ models unifying inflation with late time accelerated expansion have been considered in \cite{sergei1}, \cite{sergei2}, \cite{sergei3}, and in the fame of Horava-Lifshitz in \cite{sergei4}. Compared with the modified $f(R)$ gravity, in the present model we consider this function $f(R)$ as linear in $R$ but we generalize the model permitting extra $R_{\mu\nu}$ coupling with kinetic-like scalar term.\\
\noindent In this paper we consider the possibility of obtaining a slow rolling inflationary background with a scalar model whose kinetic term is non-minimally coupled to the scalar field and to the curvature \cite{granda}. A particular attention will be paid to the quadratic and Higgs field inflation, where the corresponding kinetic terms are non-minimally coupled to the quadratic (Higgs) field and to scalar and Ricci curvatures.
%%%%%%%%%%%%%%%%%%%%%%%%%%%%%%%%%%%%%%%%%%%%%%%%%%%%%%%%%%%%%%%%%%%%%%%%%%%%%%%%%%%%%%%%%%%%%%%%%%%%%%%%%%%%%%%%%%%%%%%%

\section{Field Equations}
We start with the following  action \cite{granda1}

\be\label{eq1}
\begin{aligned}
S=&\int d^{4}x\sqrt{-g}\Big[\frac{1}{16\pi G} R-\frac{1}{2}\partial_{\mu}\phi\partial^{\mu}\phi-\frac{1}{2} \xi R \left(F(\phi)\partial_{\mu}\phi\partial^{\mu}\phi\right) -\\ 
&\frac{1}{2} \eta R_{\mu\nu}\left(F(\phi)\partial^{\mu}\phi\partial^{\nu}\phi\right) - V(\phi)\Big] 
\end{aligned}
\ee

\noindent where the dimensionality of the coupling constants $\xi$ and $\eta$ depends on the type of function $F(\phi)$. 
Assuming the spatially-flat Friedmann-Robertson-Walker (FRW) metric
\be\label{eq2}
ds^2=-dt^2+a(t)^2\left(dr^2+r^2d\Omega^2\right),
\ee
and taking the variation of action (\ref{eq1}) with respect to the metric, in the flat FRW background we obtain the following Friedman equation (see details in \cite{granda})
\be\label{eq4}
\begin{aligned}
H^2=&\frac{\kappa^2}{3}\Big[\frac{1}{2}\dot{\phi}^2+V(\phi)+9\xi H^2F(\phi)\dot{\phi}^2+3(2\xi+\eta)\dot{H}F(\phi)\dot{\phi}^2\\
&-3(2\xi+\eta)H F(\phi)\dot{\phi}\ddot{\phi}-\frac{3}{2}(2\xi+\eta)H \frac{dF}{d\phi}\dot{\phi}^3\Big]
\end{aligned}
\ee
with the Hubble parameter $H=\dot{a}/a$.
Variating with respect to the scalar field gives rise to the equation of motion
\be\label{eq5}
\begin{aligned}
&\ddot{\phi}+3H\dot{\phi}+\frac{dV}{d\phi}+3(2\xi+\eta)\ddot{H}F(\phi)\dot{\phi}+
3(14\xi+5\eta)H\dot{H}F(\phi)\dot{\phi}\\
&+\frac{3}{2}(2\xi+\eta)\dot{H}\left(2F(\phi)\ddot{\phi}+\frac{dF}{d\phi}\dot{\phi}^2\right)+
\frac{3}{2}(4\xi+\eta)H^2\left(2F(\phi)\ddot{\phi}+\frac{dF}{d\phi}\dot{\phi}^2\right)\\
&+9(4\xi+\eta)H^3F(\phi)\dot{\phi}=0
\end{aligned}
\ee
where the first three terms correspond to the minimally coupled field. In what follows we study the cosmological consequences of this equations under some conditions that simplify the search for solutions, but nevertheless still showing interesting cosmological solutions.\\
\noindent First note that the Eqs. (\ref{eq4}) and (\ref{eq5}) significantly simplify under the restriction on $\xi$ and $\eta$ given by 
\be\label{eq1a}
\eta+2\xi=0
\ee
This restriction is equivalent to a coupling of the kinetic term to the Einstein tensor $G_{\mu\nu}$ (see \cite{capozziello1}, \cite{capozziello2}). In this case the field equations (\ref{eq4}) and (\ref{eq5}) contain only second derivatives of the metric and the scalar field. The simplified equations take the form
\be\label{eq6}
H^2=\frac{\kappa^2}{3}\left(\frac{1}{2}\dot{\phi}^2+V(\phi)+9\xi H^2F(\phi)\dot{\phi}^2\right)
\ee
and 
\be\label{eq7}
%\begin{aligned}
\ddot{\phi}+3H\dot{\phi}+\frac{dV}{d\phi}+3\xi H^2\left(2F(\phi)\ddot{\phi}+\frac{dF}{d\phi}\dot{\phi}^2\right)
+18\xi H^3F(\phi)\dot{\phi}+12\xi H\dot{H}F(\phi)\dot{\phi}=0
%\end{aligned}
\ee
In what follows we will study the Eqs. (\ref{eq6}) and (\ref{eq7}) and see how non minimal kinetic coupling can be used to obtain an inflating
background. \\

%%%%%%%%%%%%%%%%%%%%%%%%%%%%%%%%%%%%%%%%%%%%%%%%%%%%%%%%%%%%%%
\section{De Sitter and power law solutions}
By restricting the potential in the model (\ref{eq1}), we can find de Sitter and power law solutions. Let's briefly review these solutions.

Multiplying the Eq. (\ref{eq7}) by $\dot{\phi}$, and replacing the product $F(\phi)\dot{\phi}^2$ from Eq. (\ref{eq6}),
the Eq. (\ref{eq7}) reduces to first order equation with respect to the variables $\psi=\dot{\phi}^2$, $H$ and $V$, and can be written as
\be\label{eq8}
H\frac{d\psi}{dt}+\left(6H^2-\dot{H}\right)\psi+2H\frac{dV}{dt}-2\left(3H^2+\dot{H}\right)V+6\frac{H^2}{\kappa^2}\left(3H^2+2\dot{H}\right)=0
\ee
Using the fact that the variables $\psi$ and $V$ are separated, we can limit the model to the class of potentials that satisfy the restriction
\be\label{eq9}
H\frac{dV}{dt}-\left(3H^2+\dot{H}\right)V+3\frac{H^2}{\kappa^2}\left(3H^2+2\dot{H}\right)=0
\ee
then, the equation for the field $\psi$ significantly simplifies as can be seen bellow 
\be\label{eq10}
H\frac{d\psi}{dt}+\left(6H^2-\dot{H}\right)\psi=0
\ee
By considering a de Sitter solution $H=H_0$ in Eq. (\ref{eq9}), the potential takes the simple form $V=3H_0^2/\kappa^2+Ce^{3H_0t}$ (with $C$ as an integration constant), and from (\ref{eq10}) it follows for the scalar field the solution
$\phi=\frac{\dot{\phi}_0}{H_0}e^{-3H_0t}$. In terms of the scalar field, the potential is of the form 
\be\label{eq11}
V=\frac{3H_0^2}{\kappa^2}+\frac{C\dot{\phi}_0}{3H_0}\frac{1}{\phi}, 
\ee
and the coupling function from (\ref{eq6}) becomes
\be\label{eq12}
F(\phi)=-\frac{1}{9\xi H_0^2}\left(\frac{1}{2}+\frac{C\dot{\phi}_0}{9H_0^3}\frac{1}{\phi^3}\right)
\ee
In order to make the scalar coupling $\xi F>0$, we can take $\xi<0$.\\

Let's assume the solution $a(t)\propto t^p$ and replace in Eqs. (\ref{eq9}) and (\ref{eq10}). Is easy to check that a particular solution to Eq. (\ref{eq9}) is given by
\be\label{eq13}
V(t)=\frac{3p^2\left(3p-2\right)}{\kappa^2\left(3p+1\right)}\frac{1}{t^2}
\ee
which for $p>0$ may be considered as asymptotic solution at $t\rightarrow0$, and the solution to Eq. (\ref{eq10}) is 
\be\label{eq14}
\psi=\psi_0\left(\frac{t}{t_0}\right)^{-(6p+1)}
\ee
where $\psi_0=\dot{\phi}_0^2$ is the integration constant. The scalar field from Eq. (\ref{eq14}) is given by
\be\label{eq15}
\phi=\phi_0\left(\frac{t}{t_0}\right)^{-3p+1/2},\,\,\,\,\,\, \phi_0=\frac{2\dot{\phi}_0t_0}{6p-1}
\ee
With this solutions the scalar potential (\ref{eq13}), and the coupling function from (\ref{eq6}) can be written explicitly in terms of the scalar field as follows
\be\label{eq16}
V(\phi)=\frac{3p^2\left(3p-2\right)}{\left(3p+1\right)\kappa^2t_0^2}\left(\frac{\phi}{\phi_0}\right)^{\frac{4}{6p-1}}
\ee
and 
\be\label{eq17}
F(\phi)=\frac{1}{\left(\xi\kappa^2\dot{\phi}_0^2\right)\left(3p+1\right)}\left(\frac{\phi_0}{\phi}\right)^{\frac{2(6p+1)}{6p-1}}-\frac{t_0^2}{18\xi p^2}\left(\frac{\phi_0}{\phi}\right)^{\frac{4}{6p-1}}.
\ee
Note that there are not restrictions on the values of $p>1$, so the obtained solutions for $V$ and $F$ give rise to accelerated expansion.\\ 
All above shows that a power-law potential and a power law-coupling, can lead either to exponential or power-law inflation.
Despite the fact that we can obtain an inflationary background of the exponential type $a(t)\propto e^{H_0t}$ or power-law type $a(t)\propto t^p$ ($p>>1$), the above solutions permit inflation to continue forever. In the next section we study the slow-roll conditions allowing for inflationary solutions. 

%%%%%%%%%%%%%%%%%%%%%%%%%%%%%%%%%%%%%%%%%%%%%%%%%%%%%%%%%%%%%%%%%%%%%%%%%%%%%%%%%%%%%%%%%%%%%%%%%%%%%%%%%%%%%%%%%%%%%%%%%%%%%%%%%%%%%%%%%%%%
\section{The slow-roll conditions}
Let's see how the present model meets the slow-roll conditions to obtain an inflating background. 
Analyzing the equations (\ref{eq6}) and (\ref{eq7}), the slow-roll approximation in our case should be obtained through the following inequalities for a sufficiently slowly varying field $\phi$
\be\label{eq20}
6\xi H^2F(\phi)>>1,\,\,\,\,\,\,  9\xi H^2F(\phi)\dot{\phi}^2<<V(\phi),\,\,\,\, \frac{\dot{H}}{H^2}<<1.
\ee
(the election of coefficient 6 instead of 18 will be clear from bellow). The first one makes possible to neglect the free kinetic term, and the second allows to reduce the Eq. (\ref{eq6}) to
\be\label{eq21}
H^2=\frac{1}{3M_p^2}V(\phi)
\ee
where we used $\kappa^2=M_P^{-2}$, and $M_p^{-2}$ is the reduced Planck mass ($M_p=2.4\times 10^{18} Gev)$. By using the third condition from (\ref{eq20}) (which allows to neglect the last term in Eq. (\ref{eq7}) with respect to penultimate term), the evolution equation (\ref{eq7}) for the scalar field can be written in the form
\be\label{eq22}
\left(1+6\xi H^2F(\phi)\right)\ddot{\phi}+\left(1+6\xi H^2F(\phi)\right)3H\dot{\phi}+3\xi H^2\frac{dF}{d\phi}\dot{\phi}^2
+\frac{dV}{d\phi}=0
\ee
and using the first condition from (\ref{eq20}) this equation can still be reduced to
\be\label{eq23}
6\xi H^2F(\phi)\left(\ddot{\phi}+3H\dot{\phi}\right)+3\xi H^2\frac{dF}{d\phi}\dot{\phi}^2+\frac{dV}{d\phi}=0
\ee
finally, applying the slow-roll condition $\ddot{\phi}<<3H\dot{\phi}$ and the additional condition 
\be\label{eq24}
3\xi H^2\frac{dF}{d\phi}\dot{\phi}^2<<\frac{dV}{d\phi}, 
\ee
it follows that
\be\label{eq25}
3H\dot{\phi}=-\frac{1}{6\xi H^2F(\phi)}\frac{dV}{d\phi}.
\ee

Here, the role of the kinetic coupling is seen. The condition (\ref{eq24}) can be interpreted in the following way: multiplying both sides of the inequality by $F$ it follows
\be\label{eq26}
\left(3\xi H^2 F\dot{\phi}^2\right)\frac{dF}{d\phi}<<F\frac{dV}{d\phi}
\ee
noting that the factor of $\frac{dF}{d\phi}$ in this equation, undergoes the second condition in (\ref{eq20}),  is enough with fixing the approximate relation
\be\label{eq27}
V(\phi)\frac{dF}{d\phi}\approx F(\phi)\frac{dV}{d\phi},
\ee
which does not enter in conflict with the rest of the conditions, at least when we consider power-law behavior for the coupling $F$ and the potential $V$, which indeed 
are the cases we are going to discuss.\\
By other hand, in order to prevent enter in the quantum gravity domain, the constraint $R\approx 12H^2<<M_p^2/2$ should be observed, which implies for the potential
\be\label{eq28}
V(\phi)<<\frac{M_P^4}{8}
\ee
Let's apply now the above restrictions and verify their consistency. We already know that from (\ref{eq21}), the universe enters in a de Sitter phase for a flat enough potential. From (\ref{eq21}) and (\ref{eq25}) it follows that
\be\label{eq29}
\dot{\phi}\approx -\frac{M_P^2V'}{6\xi FHV}
\ee
From the second condition in (\ref{eq20}), and using (\ref{eq21}) it follows that
\be\label{eq30}
\dot{\phi}^2<<\frac{M_p^2}{3\xi F}
\ee
and replacing $\dot{\phi}$ from (\ref{eq29}) in (\ref{eq30}) (taking into account (\ref{eq21}))
\be\label{eq31}
\frac{V^3}{V'^2}>>\frac{M_P^4}{4\xi F}
\ee
This last relation is consistent with the definition of the slow-roll parameter $\epsilon$ as
\be\label{eq31a}
\epsilon=-\frac{\dot{H}}{H^2}\approx \frac{M_P^4}{4\xi F}\frac{V'^2}{V^3},
\ee
where we have used (\ref{eq21}) and (\ref{eq25}). We can also check the consistency conditions for the slow-roll, by taking the derivative of $\dot{\phi}$  in Eq. (\ref{eq25}), which gives
\be\label{eq31b}
\ddot{\phi}=\left(\frac{V'H'}{6\xi FH^4}-\frac{V''}{18\xi FH^3}+\frac{V'F'}{18\xi F^2 H^3}\right)\dot{\phi}
\ee
and comparing with the condition $\ddot{\phi}<<3H\dot{\phi}$, gives consistency provided that
\be\label{eq31c}
\frac{V'H'}{18\xi FH^5}=\frac{M_P^4V'^2}{4\xi FV^3}=\epsilon<<1,\,\,\,\,\, \left|\frac{V''}{54\xi F H^4}\right|= \left|\frac{M_P^4 V''}{6\xi F V^2}\right|=\eta<<1
\ee
and the last term in Eq. (\ref{eq31b}), can be reduced by using (\ref{eq27}) (replacing $F'$), to  $2\epsilon/3$.\\
Note that by making $F=1$, $\xi=w^2/2$, and using the Higgs potential $V=\frac{\lambda}{4}\phi^4$, the results discussed in \cite{germani} can be reproduced (in general the Higgs potential is of the form $V=V\frac{\lambda}{4}(\phi^2-v^2)^2$, but here is assumed that during inflation $\phi>>v$). 
Applying to the Higgs potential and taking the coupling function of the form $F(\phi)=1/\phi^2$ (see \cite{granda1}), the restriction (\ref{eq31})) gives
\be\label{eq32}
\phi^4>>\frac{16M_P^4}{\lambda\xi}
\ee
where $\xi$ is dimensionless now. But from (\ref{eq28}) it follows that 
\be\label{eq33}
\phi^4<<\phi_c^4\equiv\frac{M_P^4}{2\lambda}
\ee
and therefore, the coupling $\xi$ should satisfy $\xi>>32$ (the critical value $\phi_c$ is the limit beyond which the quantum corrections become important). The number of e-folds during inflation are given by
\be\label{eq34}
N=\int^{\phi_{0}}_{\phi_{end}}\frac{H}{\dot{\phi}}d\phi\approx \int^{\phi_{0}}_{\phi_{end}}\frac{2\xi FV^2}{M_P^4V'}\approx\frac{\xi\lambda\phi_0^4}{32M_P^4}
\ee
\noindent where we have used (\ref{eq29}) and assumed that the scalar field at the end of inflation is a small fraction of $\phi_0$. In terms of the critical value $\phi_c$ (\ref{eq33}) and considering $\lambda\sim 0.2$ and $N\sim 60$ we get $(\phi_0/\phi_c)^4\approx 3.8\times10^3/\xi$. If we take for example $\phi_0/\phi_c\sim 5\times10^{-4}$ (i.e. $\phi_0\sim 6.25\times 10^{-4}M_P$ according to (\ref{eq33})), then $\xi\sim 1.5\times 10^{18}$. This value is not over sized as the coupling is given by the product $\xi F\approx \xi/\phi_0^2\sim 0.68 (10^3Tev)^{-2}$.\\
\noindent In a more general case of power-law potential $V=\lambda \phi^n/n$ and maintaining the same coupling, the restrictions (\ref{eq28}) and (\ref{eq31}) become
\be\label{eq35}
\phi^n<<\frac{nM_P^4}{8\lambda},\,\,\,\,\,\, \phi^n>>\frac{n^3M_P^4}{4\xi\lambda}
\ee
which are consistent for $\xi>>2n^2$. Applied to the $n=2$ case, and making $\lambda=m^2$ the number of $N$-folds from (\ref{eq34})) gives the result
\be\label{eq36}
N\approx\frac{\xi m^2}{4M_P^4}\phi_0^2=\frac{\xi}{16}\left(\frac{\phi_0}{\phi}\right)^2
\ee
where we have used (\ref{eq28}) to define the critical field $\phi_0$ from the restriction 
\be\label{eq37}
\phi^2<<\phi_c^2\equiv \frac{M_P^4}{4m^2}.
\ee
\noindent For $N\approx 60$, we get $(\phi_0/\phi_c)^2\approx 10^3/\xi$. Taking for instance $\phi_0/\phi_c\sim 10^{-9}$, then $\xi\approx 10^{21}$, and giving a value for $m\approx 5\times10^{-6}M_P$, it is obtained $\phi_0\approx 10^{-4}M_P$ and the coupling $\xi F$ becomes of the order of $\xi F\sim 1.8(10 Tev)^{-2}$. To have an appreciation of the scalar field after the conditions for slow-rolling have expired,  the condition $\epsilon\sim 1$ can be applied. For the power-law potential of the form $V=(\lambda/n)M_P^{4-n}\phi^n$, and from (\ref{eq31a}), the condition $\epsilon\sim 1$ gives 
\be\label{eq38}
\phi_{end}\approx \left(\frac{n^3}{4\xi\lambda}\right)^{1/n}M_p,
\ee
which for the quadratic potential ($n=2$ and $\lambda M_P^2=m^2$) gives $\phi_{end}\approx(2/\xi)^{1/2}M_P^2/m$. For the the values we considered above for $m$ and $\xi$, it is found $\phi_{end}\approx 10^{-5}M_P$. Applied to the Higgs potential $n=4$, $\phi_{end}\approx 2M_P/(\xi\lambda)^{1/4}\approx 10^{-5}M_P$ (for $\lambda\sim 0.2$ and $\xi\sim 10^{20}$). One interesting aspect of this model, is that unlike the simplest single-field models of inflation (that require super-Planckian field values, $\phi\geq M_p$), in the present model one does not need a large scalar field to develop the necessary number of e-foldings of inflation.
%%%%%%%%%%%%%%%%%%%%%%%%%%%%%%%%%%%%%%%%%%%%%%%%%%%%%%%%%%%%%%%%%%%%%%%%%%%%%%%%%%%%%%%%%%%%%%%%%%%%%%%%%%%%%%%%%%%%%%%%%%%%%%%%%
%%%%%%%%%%%%%%%%%%%%%%%%%%%%%%%%%%%%%%%%%%%%%%%%%%%%%%%%%%%%%%%%%%%%%%%%%%%%%%%%%%%%%%%%%%%%%%%%%%%%%%%%%%%%%%%%%%%%%%%%%%%%%%%%%
\section{Qualitative analysis of predictions}
In this section we present some qualitative arguments that favor the present model as phenomenologically viable inflationary model.
Most of the estimates are rough and the results are given for the choice of the parameters used in previous section, but this not rule out another choice of parameters that might fit the observations.
The parametrization of slow roll models by $\epsilon$ and $\eta$ enables one to explore some predictive consequenses for observables without assuming specific model (at least in firs order approximation). Each model predicts the slow roll parameters, and in our case some characteristics of the kinetic coupling are present in the slow roll parameters, which allow to predict the observables. As will be seen bellow, the smallness of the slow roll parameters justify the approximation used here to express the scalar spectral index though $\epsilon$ and $\eta$.\\
From (\ref{eq31c}) it follows for the quadratic potential $V=1/2 m^2\phi^2$
\be\label{eq39}
\epsilon=\frac{M_p^4 V'^2}{4\xi F V^3}=\frac{8M_p^4}{4\xi F m^2\phi_0^4}
\ee
where $\phi_0$ is the value of the scalar field during inflation.
Using $\xi F\approx 1.8(10 Tev)^{-2}$, $m\approx 5\times 10^{-6}M_P$ and $\phi_0\approx 10^{-4}M_P$, gives $\epsilon \approx 8\times 10^{-3}$.\\
For the $\eta$ parameter it follows 
\be\label{eq40}
\eta=\frac{M_p^4 V''}{6\xi F V^2}=\frac{2M_p^4}{3\xi F m^2 \phi^4}
\ee
replacing the above values, gives $\eta\approx 2.7\times 10^{-3}$.\\
To first order and using the standard slow roll analysis, for the case under consideration the results are modified by the kinetic coupling, allowing for observationally distinct 
signatures, which are encoded at least in first approximation, in the dependence of the slow roll parameters on the potential and the coupling as given by (\ref{eq31c}). To lowest order in the slow roll parameters, we will consider the scalar spectral index in the form (see \cite{liddle} for the definition and behavior of the scalar power spectra)
\be\label{eq41}
n_s=1-6\epsilon+2\eta
\ee
Although this approximation is based on the non minimally coupled scalar field, the inequality: $9\xi H^2F\dot{\phi}^2<<1$ (valid during inflation, see (\ref{eq20})) allows us to use this approximation as follows from (\ref{eq6}). For the quadratic potential, from the above values it is obtained $n_s\approx 0.96$. For the Higgs potential, and using the values: $\xi F\sim 0.68 (10^3 Tev)^{-2}$, $\phi_0\sim 6.25\times 10^{-4}M_P$ and $\lambda \sim 0.2$, it  is obtained
\be\label{eq42}
\epsilon=\frac{16 M_p^4}{\lambda \xi F \phi^6}\approx 3.4\times 10^{-3}
\ee
and
\be\label{eq43}
\eta=\frac{8 M_p^4}{\lambda \xi F \phi^6}=\frac{\epsilon}{2}\approx 1.7\times 10^{-3}
\ee
and the scalar spectral index takes the value $n_s=1-5\epsilon\approx 0.98$ which is very close to the scale invariance value ($n_s=1$) \cite{liddle}, \cite{peiris}. So the above choice of the fields and parameters during inflation (that give an appropriate number of e-foldings $N\sim 60$) gives an scalar spectral index in the observational range.\\
After the inflation ends, the scalar field moves towards its vev and typically oscillates before settling down. To reach this conclusion in the present model, let's write Eq. (\ref{eq23}) as 
%(the first of the restrictions (\ref{eq20}) still taking place after inflation).
\be\label{eq44}
\ddot{\phi}+3H\dot{\phi}+\frac{1}{2F}\frac{dF}{d\phi}\dot{\phi}^2+\frac{V'}{6\xi H^2 F}=0
\ee
writing the last term in Eq. (\ref{eq44}) as $V'_{eff}$, and the third term as $1/2 (d(\log{F})/dt)\dot{\phi}=\Gamma_{\phi}\dot{\phi}$, then the Eq. (\ref{eq44}) can be rewritten as
\be\label{eq45}
\ddot{\phi}+3H\dot{\phi}+\Gamma_{\phi}\dot{\phi}+V'_{eff}=0
\ee
similar equation is used to describe the semi classical evolution of decaying scalar field (see \cite{steinhardt1}, \cite{linde5}). If we initially neglect the third term, then the scalar field just after inflation begins executing oscillations, which are damped by the ``friction'' term $3H\dot{\phi}$. Once the amplitude of oscillations start decreasing at enough rate, the energy density of the scalar field begins to be transferred  to other matter fields and the reheating takes place (see \cite{linde5})). Concerning the third term in Eq. (\ref{eq45}), it plays a role similar to the second term. The justification for ignoring this term initially, is that we considered that the time magnitude associated with this term (i.e. $\Gamma_{\phi}^{-1}$, which could be interpreted as the life time of the Higgs boson associated with $\phi$) is much greater than $H^{-1}$. Depending on the magnitude of this term it would influence the effect of the ``friction'' term and the transition to the reheating regime becomes more or less efficient, but the description of the reheating process requires the coupling of the inflaton $\phi$ to other matter fields.  
This term could also be interpreted as describing the decay of the inflaton into its own modes due to the gravitational and self interactions.

\section{Discussion}
%%%%%%%%%%%%%%%%%%%%%%%%%%%%%%%%%%%%%%%%%%%%%%%%%%%%%%%%%%%%%%%%%%%%%%%%%%%%%%%%%%%%%%%%%%%%%%
Despite the success of the inflationary paradigm in resolving the problems
of the standard big bang model and in providing a mechanism for the formation of structures
in the universe, there is no universally accepted model for inflation, and many different inflationary scenarios have been proposed. 
Moreover, it has not been possible to unambiguously identify the inflaton with any known field from a particle physics theory.
A comparison of the inflationary models with observations has been made possible in
recent years by the discovery of anisotropies in the cosmic microwave background \cite{komatsu}. In the present paper, we study a scalar field model with
non-minimal kinetic coupling to the scalar field, and to scalar and Ricci curvature terms, and considered quadratic and Higgs type potentials. For power-law potentials in terms of the scalar field, the model presents exponential and power-law inflationary solutions without slow-roll conditions. It was also shown that the model produces
a successful slow rolling inflation, without exceeding the quantum gravity bound. For the quadratic potential we can get enough inflation provided
$m$ is small enough as required form observational limits on the size of density perturbations, and the value of the scalar field at the end of inflation
has an appropriate value for the standard model of particle theory. In the case of the Higgs potential, we have used for the self coupling the value $\lambda\sim 0.2$ (which is in the range set by current high energy physics experiments \cite{data}) and obtained the appropriate number of e-folds ($N\sim 60$) for small enough inflaton field, and at the end of inflation the field may have also an appropriate value according to the high energy particle theory. Thus, one important property of the inflation under this kinetic coupling is that the values of the fields at the end of inflation allows to recover the connection with particle physics. 
We have also presented some qualitative calculations that give estimates about the scalar spectral index, showing that the model is phenomenologically viable as inflationary model. Its worth mention that the classical equation of motion for the scalar field contains a decaying-like term, usually introduced phenomenologically to account for the reheating or obtained in a semi classical approach \cite{steinhardt1}, \cite{linde5}. In our case this term would make the reheating process more or less efficient and would describe the decay of the inflaton field into its own lighter modes. All these observations lead us to conclude that the implications of this model deserve a deeply study.

\section*{Acknowledgments}
This work was supported by in part by the SENECA Foundation (Spain), program PCTRM 2007-2010.

\end{document}